\documentclass[iop]{emulateapj}

\usepackage{psfig}
\usepackage{graphicx}
\usepackage{natbib}
\usepackage[english]{babel}  
\usepackage{txfonts}

\shorttitle{Spatial distribution of the ejecta in Tycho's SNR}
\shortauthors{Miceli et al.}

\begin{document}

\title{Spatial distribution of X-ray emitting ejecta in Tycho's SNR: indications of shocked Titanium}

\author{M. Miceli\altaffilmark{1,2}, S. Sciortino\altaffilmark{2}, E. Troja\altaffilmark{3,4}, S. Orlando\altaffilmark{2}}

\altaffiltext{1}{Dipartimento di Fisica \& Chimica, Universit\`a di Palermo, Piazza del Parlamento 1, 90134 Palermo, Italy \email{miceli@astropa.unipa.it}}
\altaffiltext{2}{INAF-Osservatorio Astronomico di Palermo, Piazza del Parlamento 1, 90134 Palermo, Italy} 
\altaffiltext{3}{NASA, Goddard Space Flight Center, Greenbelt, MD 20771, USA}
\altaffiltext{4}{Department of Physics and Department of Astronomy, University of Maryland, College Park, MD 20742, USA}

\begin{abstract}
Young supernova remnants show a characteristic ejecta-dominated X-ray emission that allows us to probe the products of the explosive nucleosynthesis processes and to ascertain important information about the physics of the supernova explosions.
Hard X-ray observations have recently revealed the radioactive decay lines of $^{44}$Ti at $\sim67.9$ keV and $\sim78.4$ keV in the Tycho's SNR. We here analyze the set of \emph{XMM-Newton} archive observations of the Tycho's SNR. We produce equivalent width maps of the Fe K and Ca XIX emission lines and find indications for a stratification of the abundances of these elements and significant anisotropies. We then perform a spatially resolved spectral analysis by identifying five different regions characterized by high/low values of the Fe K equivalent width. We find that the spatial distribution of the Fe K emission is correlated with that of the Cr XXII. We also detect the Ti K-line complex in the spectra extracted from the two regions with the highest values of the Fe and Cr equivalent widths. The Ti line emissions remains undetected in regions where the Fe and Cr equivalent widths are low. Our results indicate that the post-shock Ti is spatially co-located with other iron-peak nuclei 
in Tycho's SNR, in agreement with the predictions of multi-D models of Type Ia supernovae. 
\end{abstract}

\keywords{X-rays: ISM --- ISM: supernova remnants --- ISM: individual object: Tycho's SNR}

\section{Introduction}
\label{Introduction}

Supernova remnants (SNRs) govern the physical and chemical evolution of our Galaxy. 
An exploding star releases $\sim10^{51}$ erg of kinetic energy through some solar masses of ejecta that expand supersonically and drive powerful shocks back and forth in the ambient medium and ejecta themselves.
The X-ray emission from young SNRs is a powerful diagnostic tool to study the imprint of the supernova explosion in the evolution of the remnant. The X-ray emission of young SNRs is, in fact, ejecta-dominated, being mainly associated with the metal-rich material expelled in the supernova explosion and heated up to X-ray emitting temperatures by the interaction with the reverse shock.
The ejecta carry information about the explosive nucleosynthesis processes and can "keep memory" of the physics of the explosion itself (e. g., \citealt{bhc08}, \citealt{mdb06}). 

In particular, the iron-group elements (e. g., Cr, Mn, Ti, together with Fe and Ni) are synthesized in the inner layers of the exploding star and can provide important information on the progenitor \citep{bbh08,ybp14}.
It has been shown that there is a correlation in the centroids of X-ray line complexes of Cr, Mn, and Fe in a large number of SNRs, including Kepler, W49B, N103B, Tycho, G344.7-0.1, and Cas A \citep{ytl13}. This result seems to suggest that these elements are spatially co-located in the explosions in this large sample of SNRs, that includes both core-collapse and Type Ia SNRs.
On the other hand, it has been recently shown that the spatial distribution of $^{44}$Ti is significantly different from that of Fe in the Cassiopeia A SNR \citep{ghb14}. 
However, the radioactive emission trace the whole amount of Ti, concentrated in the unshocked interior of the remnant, while the X-ray emission from Fe originates only from the ejecta shocked by the reverse shock. Therefore, the different morphologies may be due to the fact that we only observe a small fraction of Fe in X-rays.

The radioactive hard X-ray signature of $^{44}$Ti (whose radioactive decay lines are at 67.86 keV and 78.36 keV) has been recently observed in Tycho's SNR through the analysis of $Swift/BAT$ \citep{tsl14} and $INTEGRAL$ \citep{wl14} observations. 
Tycho's SNR is the remnant of a Type Ia SN explosion \citep{bbh06,ktu08} occurred in 1572 AD and presents clear signatures of efficient particle acceleration (see, e.g., \citealt{ehb11,beo11,mc12,sle14}). Besides regions characterized by strong synchrotron X-ray emission, its thermal X-ray radiation is dominated by the ejecta \citep{chb07}.

In Type Ia SNRs, delayed-detonation models (e.~g., \citealt{gko05}) predict that Fe-group elements are located in the inner parts of the ejecta profile, surrounded by intermediate-mass elements (e. g., Si, S, and Ca). More recently, three-dimensional delayed-detonation models developed by \citet{scr13} revealed the details of the element stratification, by showing that Fe-group elements can have velocities higher than those of $^{56}$Ni (which, after the decay, produces the bulk of Fe-rich ejecta), but still lower than those of the intermediate-mass elements, which are then expected to expand in an outer shell. 
Three-dimensional deflagration models (e.~g., \citealt{rh05}) generally suggest a more efficient mixing in the abundances distribution than classical 1-D deflagration models \citep{nty84}. 

We here take advantage of the deep set of \emph{XMM-Newton} archive observations of the Tycho's SNR to study the spatial distribution of the heavy elements in the shocked ejecta. We also look for the Ti K-emission line complex at $\sim4.9$ keV, to check whether the spatial distribution of the shocked Ti somehow correlates with that of the shocked Fe-rich ejecta and of the other Fe-group elements. The Ti K-line has not been observed yet in any SNRs: only some indications of the Ti He$\alpha$ emission line in the $ASCA$ spectrum of W49B has been reported \citep{hph00}, though it was not confirmed by the subsequent \emph{XMM-Newton} observations \citep{mdb06}.

\section{Results}
\label{results}

We analyzed the archive \emph{XMM-Newton} EPIC observations 0096210101, 0310590101, 0310590201, 0412380101, 0412380201, 0412380301, 0412380401, 0511180101, all having pointing coordinates $\alpha_{J2000}=00^h25^m22.0^s$, $\delta_{J2000}=+64^{\circ}08'24.0"$. 
Data were processed with the Science Analysis System (SAS V12). We selected events with PATTERN$\le12$ for the MOS cameras, PATTERN$\le$4 for the pn camera, and FLAG=0 for both. 
We inspected all the light curves and adopted the ESPFILT task (which is based on a sigma-clipping algorithm) to remove high background periods from the event lists, thus filtering out the contribution of flaring background associated with soft protons. We obtained a total screened exposure time for the pn observations of $124.9$ ks.
Event files were processed with the EVIGWEIGHT task to correct for vignetting effects. Images were produced by adopting the procedure described in \citet{mdb06} (see their Sect. 2).
Spectral analysis was performed in the energy band $3.6-6.7$ keV using XSPEC V12.8.2. This band was chosen so as to include Ti K-line and the characteristic X-ray line emission from other Fe-group elements (Cr, Mn, and Fe). We also carefully modelled the bright Ca emission to study the intermediate mass elements and to account for faint Ca XIX and Ca XX transitions which have energy of $\sim4.9$ keV and may contribute to the flux in the energy band where we expect to find the Ti K lines. The pn spectra of all the different observations were fitted simultaneously. 
For each spectrum, we subtracted a background spectrum extracted from a nearby region immediately outside of the SNR shell and we verified that the best-fit values do not depend significantly on the choice of the background region.
In all the fittings, the column density of the interstellar absorption was fixed to $N_H=7\times10^{21}$ cm$^{-2}$ in agreement with \citet{chb07}. Small local variations of the $N_H$ can be present across the remnant (\citealt{sle14}), but we do not expect any significant effects on our results, given the relatively hard energy band considered here.

\subsection{Image Analysis}
To trace the spatial distribution of the chemical abundances in the shocked ejecta, we produced equivalent width (EW) maps. X-ray line and continuum emission both scale with the plasma emission measure, and EW maps allow us to disentangle higher element abundances from higher emission measure. Though the EW of an emission line increases with the element abundance, it also varies with the plasma temperature and ionization age (e.g, \citealt{vb99}). Therefore, EW maps provide indications of the distribution of the abundances, but do not give quantitative information and need to be tested with spatially resolved spectral analysis (as we do in Sect. \ref{spectra}).

To identify the Fe-group elements, we produced the equivalent width (EW) map for the Fe K line complex, which, thanks to the high statistics of the \emph{XMM-Newton} data, has a much higher signal-to-noise ratio than that presented in \citet{hdh02}. We also produce, for the first time, the EW map of the Ca XIX emission lines and, for comparison, of the Si XIII lines. To produce these maps we divided the continuum-subtracted line images (in the $1.65-2.05$ keV, $3.6-4.05$ keV, and $6.1-6.7$ keV energy bands for Si, Ca and Fe, respectively) by the corresponding underlying continuum. The underlying continuum was estimated by modelling the global\footnote{To exclude the contribution of the synchrotron emitting limbs, we extracted the spectrum from a circular region slightly smaller than the shell.} pn spectrum of the remnant in a continuum band adjacent to the line emission with a phenomenological power-law model. In particular, we considered the $4.4-6.1$ keV band for Ca and Fe (with 
best-fit photon index $\Gamma=2.6$), and the $1.47-1.65$ keV band for the Si line (in this case, with a best-fit $\Gamma=1.65$). 
Though spatial variations of the spectral slope over the SNR may locally affect these maps, our spatially resolved spectral analysis shows that the EW maps provide reliable results (see Sect. \ref{spectra}). 

\begin{figure*}[tb]
  \centerline{\hbox{     
      \psfig{figure=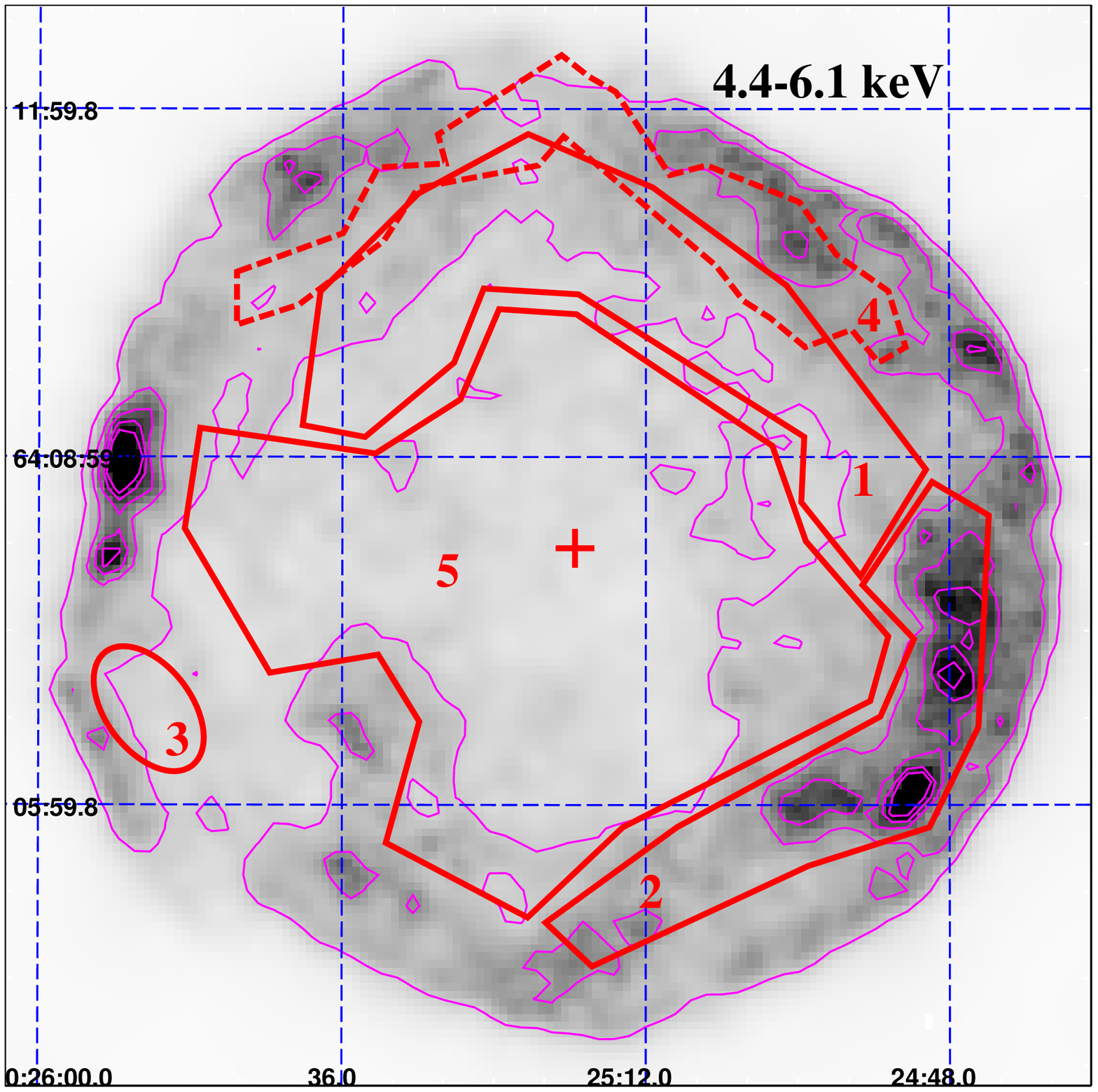,width=\columnwidth}
      \psfig{figure=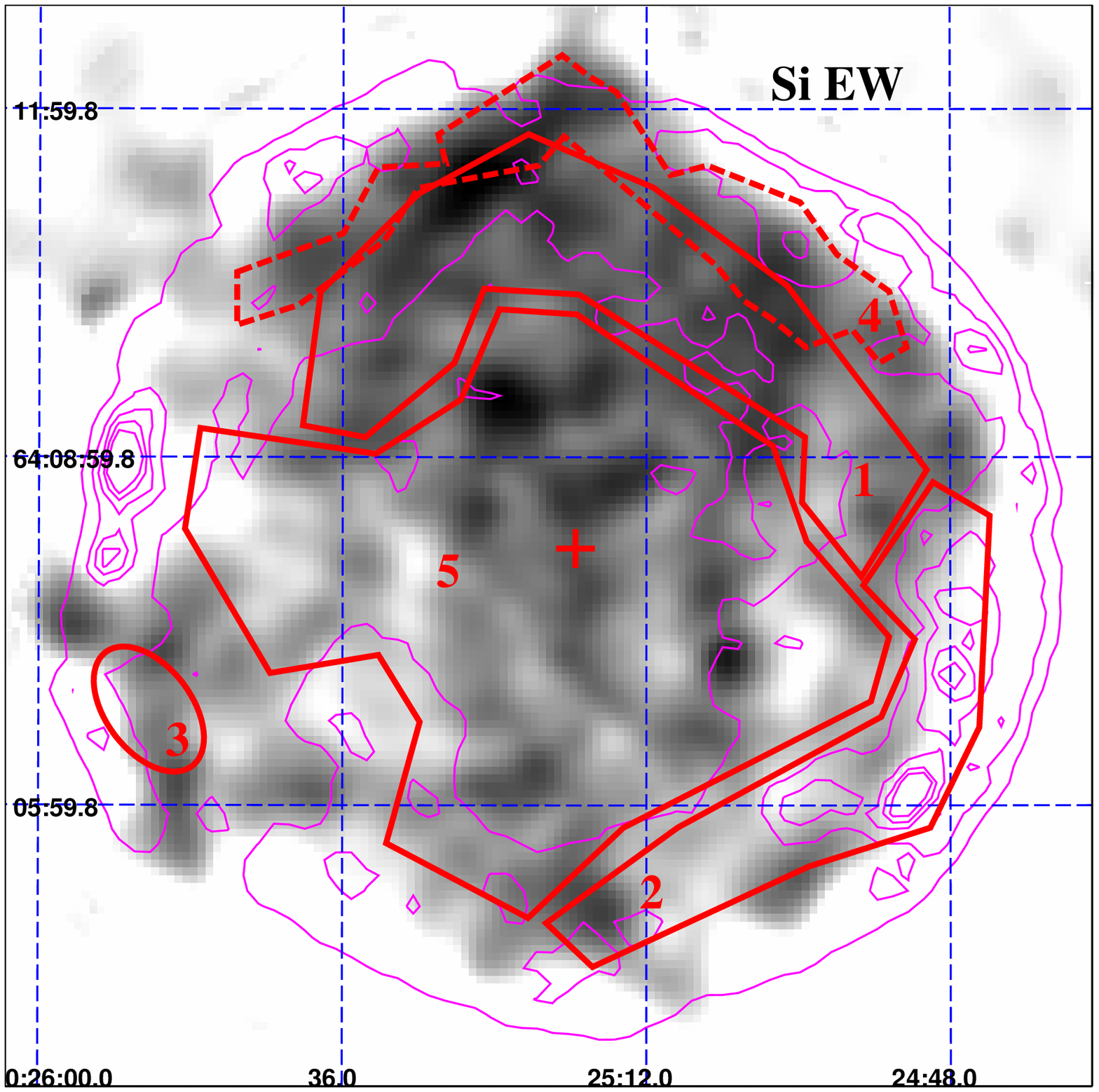,width=\columnwidth}}}
  \centerline{\hbox{     
      \psfig{figure=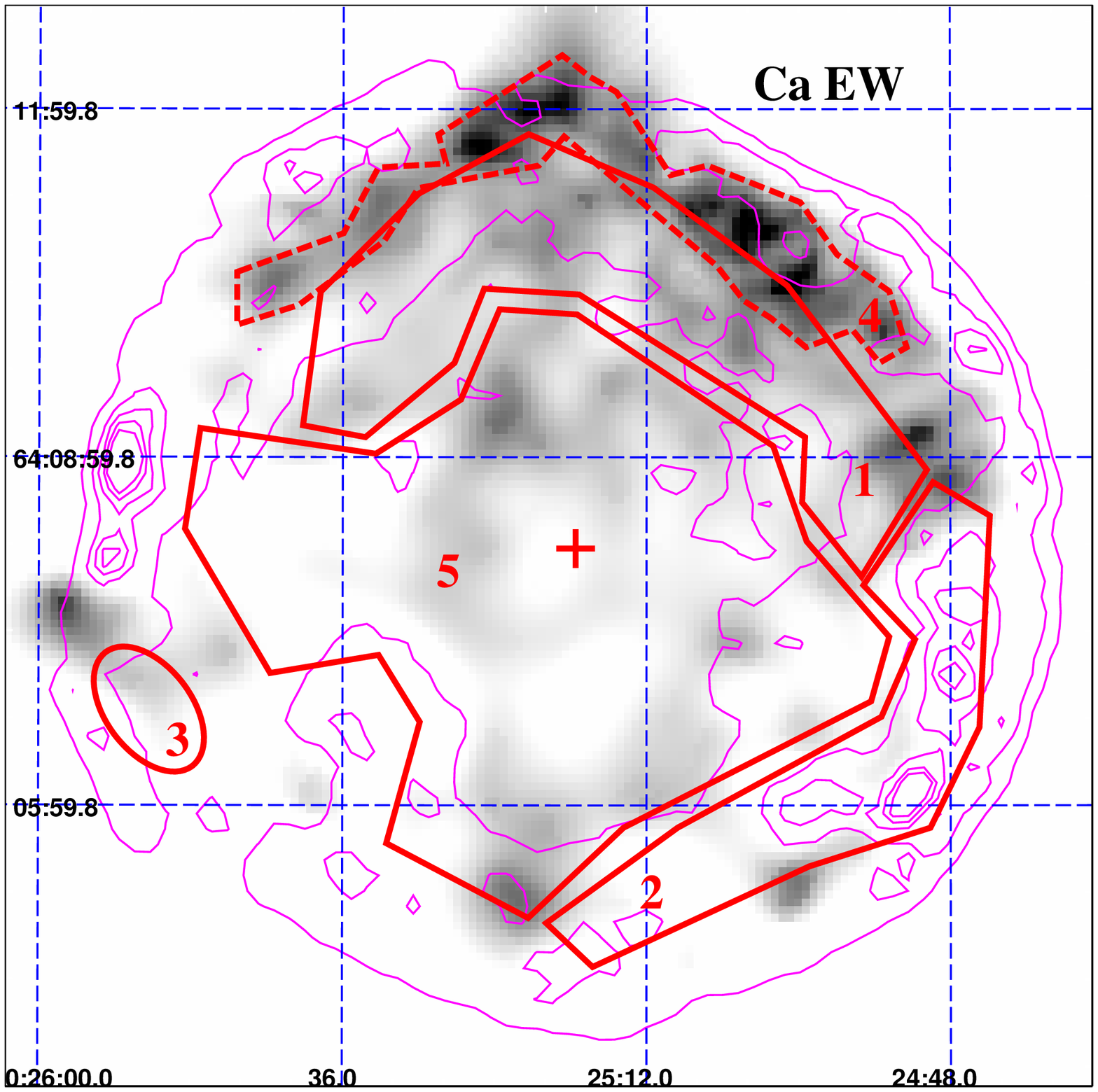,width=\columnwidth}
      \psfig{figure=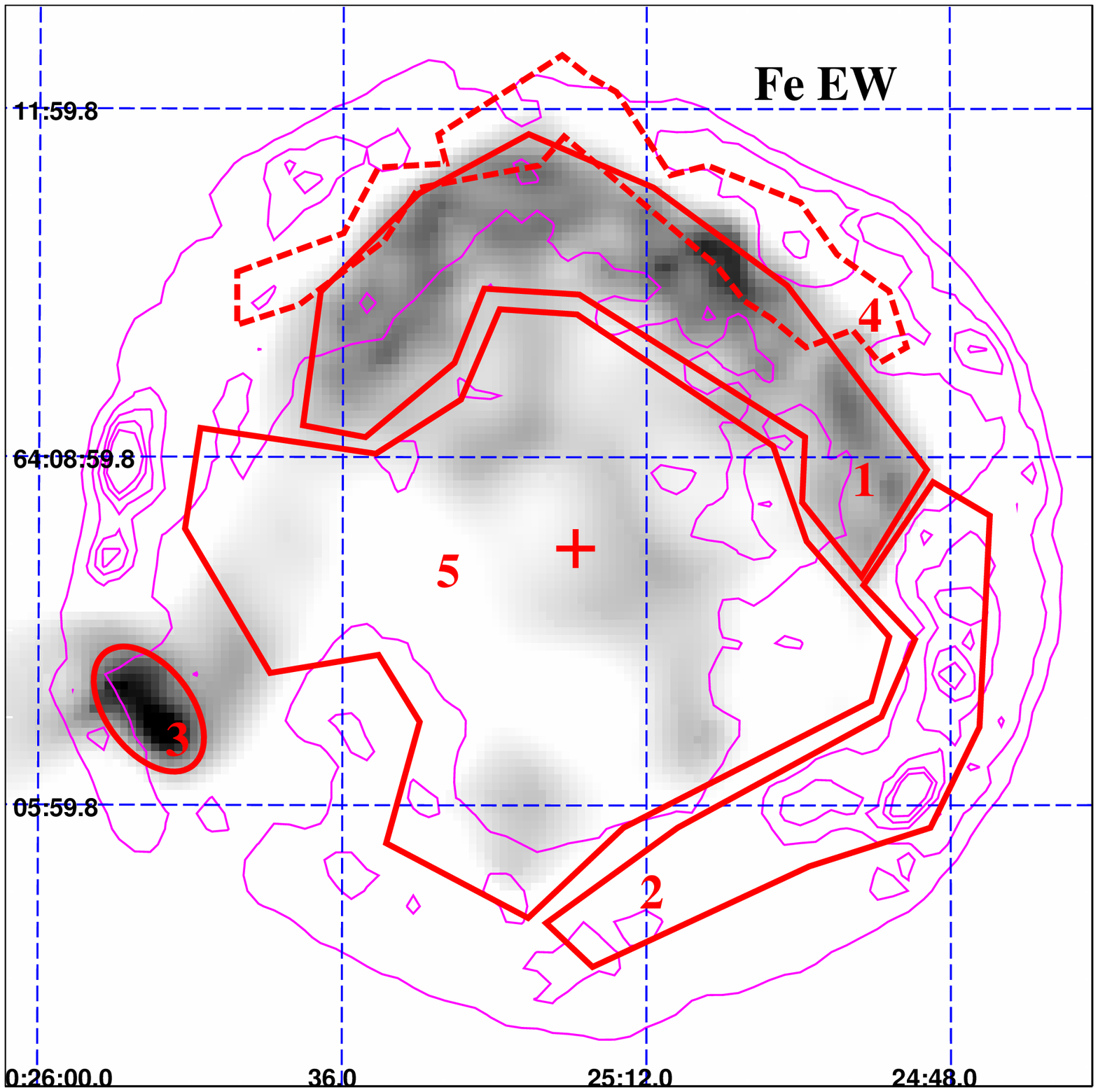,width=\columnwidth}}}      
 \caption{\emph{Upper left panel:} EPIC count-rate images (MOS and pn mosaic) of Tycho's SNR in the continuum band  $4.4-6,1$ keV. \emph{Upper right panel:} Si EW map obtained in the $1.65-2.05$ keV keV band. \emph{Lower left panel:} Ca EW map obtained in the $3.6-4.05$ keV band. \emph{Lower right panel:} the Fe K EW map obtained in the $6.1-6.7$ keV band. The bin size is $4''$ and the images are all adaptively smoothed to a signal-to-noise ratio $R=10$, but the Si EW map , where $R=25$. We superimposed the regions selected for the spectral analysis (in red) together with the contour levels of the continuum image at $10\%,~20\%,~30\%,~40\%$, and$50\%$ of the maximum (in magenta). The cross marks the position chosen as the center of the shell. North is up and East is to the left.}
 \label{fig:map}
 \end{figure*}

\begin{figure}[tb]
  \centerline{\hbox{     
      \psfig{figure=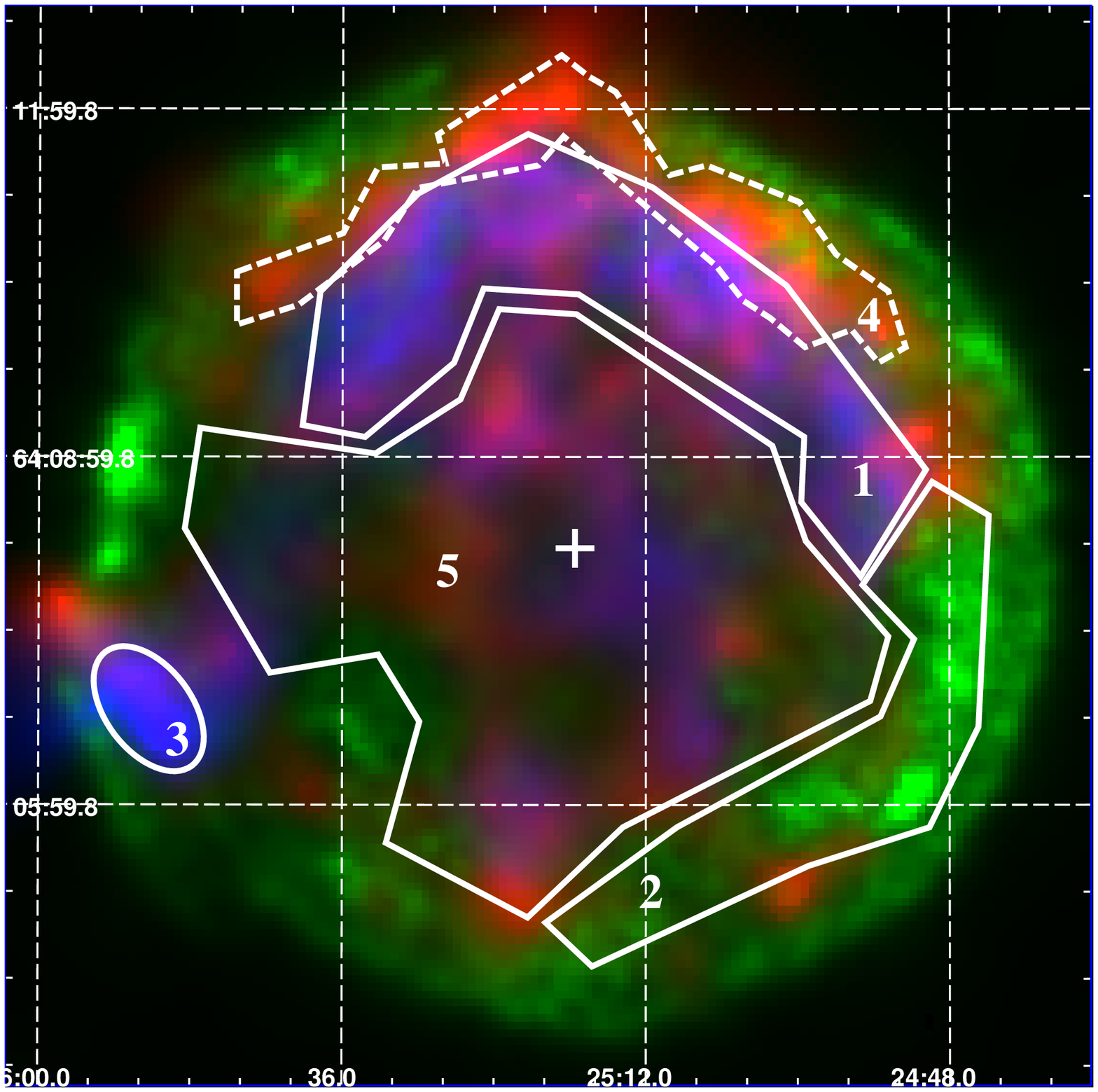,width=\columnwidth}}}
 \caption{Color-composite image showing the EPIC count-rate images in the $4.4-6,1$ keV band (\emph{green}), together with the Ca EW map (in $red$), and the Fe K EW map (in $blue$). The regions selected for the spectral analysis are superimposed.}
 \label{fig:rgb}
 \end{figure}

Fig. \ref{fig:map} shows the continuum ($4.4-6.1$ keV) count-rate image of the Tycho's SNR, together with the Si, Ca, and Fe EW maps. While the Si EW is fairly distributed over the whole remnant, (in agreement with the Si EW map obtained with $Chandra$ by \citealt{hdh02}), our Fe and Ca EW maps reveal strong anisotropies, with high values at North and relatively low values in the center and elsewhere in the rim. It is not easy to ascertain the origin of these inhomogeneities, which may be intrinsic or resulting from the interaction of the remnant with the ambient medium. In fact, Tycho's SNR evolves in a structured environment (see, e.g., \citealt{wbg13,cks13}) that may produce anisotropies and corrugations in the reverse shock front. On the other hand, these anisotropies are not present in the Si-rich ejecta, therefore the observed Ca and Fe inhomogeneities may be intrinsic in the ejecta structure. 

The Ca XIX EW presents a bright arc immediately behind the northern border of the shell (region 4 in Fig. \ref{fig:map}). A large region characterized by strong Fe EW is clearly present behind this bright Ca arc, closer to the center of the remnant (region 1 in Fig. \ref{fig:map}). Fig.\ref{fig:rgb} shows the comparison between the Ca and Fe EWs (and the continuum emission) to highlight the differences in the distribution of these elements.
The spatial distribution of the Si, Ca, and Fe EWs is suggestive of a possible stratification of the SN ejecta, with Fe confined within the inner region and lower-Z elements forming an outer envelope.
An indirect spectroscopic indication of a stratification between Fe-rich ejecta and lighter elements was proposed on the basis of the analysis of the global spectrum of the remnant (e.g., \citealt{hhp98}) and of the spectrum extracted from a large region in the eastern part of the shell (\citealt{bbh06}). But with our EW map (together with the spatially resolved spectral analysis presented in Sect. \ref{spectra}), we can directly separate out the different elements spatially, thus revealing that, besides the element stratification, there are also strong anisotropies in the Ca and Fe distributions. 

A stratification in the ejecta abundances is predicted by 3-D simulations of delayed-detonation Type Ia SNe, which show a turbulent inner region characterized by iron-group elements, surrounded by a smooth distribution of intermediate-mass elements (including Ca) in the outer layers of the ejecta \citep{krw09}. This stratification between Fe and Ca is also observed in Type Ia SNe (e.g., \citealt{tms11}). As for the inhomogeneities in the Ca map, the light-echo spectrum of Tycho's SN has revealed a high velocity component in the Ca-rich ejecta \citep{ktu08}. Also, the presence of fast Ca-rich knots deduced by \citet{ktu08}, may explain why the Ca EW presents a much higher level of anisotropy than the Si EW. Our results then suggest that the remnant has kept memory of the 
pristine ejecta distribution.

We also confirm the presence of the bright Fe-rich eastern knot at south-east \citep{vgh95}. Our map shows that this knot (region 3  in Fig. \ref{fig:map}) has the highest Fe K EW observed in the whole remnant (see also Sect. \ref{spectra} and  Fig. \ref{fig:ew}). 
We find that the projected distance between this knot and the approximate center of the remnant (indicated by a yellow cross in Fig. \ref{fig:map}) is $30\%$ higher than that of region 1 (i. e. of the ``smooth" distribution of shocked Fe-rich ejecta). This difference can be explained by considering the Fe-rich knot as a moderately overdense shrapnel (a density inhomogeneity in the ejecta profile) that moves beyond the Fe-rich ejecta shell, as modelled by \citet{mor13}. Ejecta shrapnels are common in core-collapse SNRs, but localized clumps of ejecta have been observed also in Type Ia SNe and their evolution has been modelled with hydrodynamic simulations (\citealt{wbh03,wbh06}, \citealt{obm12}), so we can argue that the eastern knot traces an ejecta clump originated in the inner layers of the exploding progenitor star.
 
\subsection{Spectral Analysis}
\label{spectra}
We performed a spatially resolved spectral analysis by selecting 5 regions defined on the basis of the Ca and Fe EW maps and shown in Fig. \ref{fig:map} and in Fig. \ref{fig:rgb}. Region 1 and 3 are very bright in the Fe K EW map and are expected to be Fe-rich; region 2 traces a part of the shell characterized by a bright continuum and low EW for both Ca and Fe; region 4 is where the Ca EW is the highest, while region 5 shows low values of the continuum and of the EW of Fe and Ca.
 \begin{figure}[tb]
  \centerline{\hbox{     
      \psfig{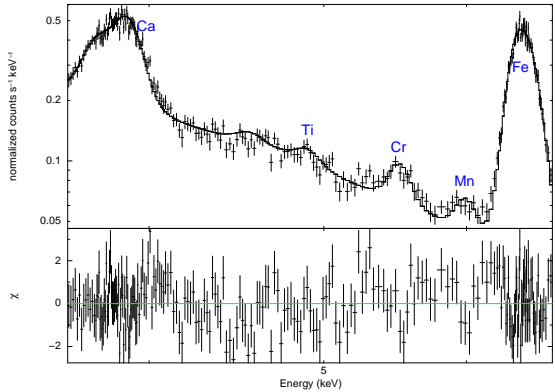}}}
 \caption{Summed pn spectrum  extracted from region 1 of Fig. \ref{fig:map}, together with its best fit model and residuals. The elements contributing to the emission lines described in the text are labelled in blue.}
 \label{fig:spec}
 \end{figure}

Figure \ref{fig:spec} shows the pn spectrum of region 1 obtained by summing all the observations (the analysis has been performed simultaneously on all the different spectra, we show the summed spectrum only for visibility reasons). We found that, in this spectrum and in all the other spectra, both the Ca and the Fe line complexes are quite broadened, with the Ca clearly showing a double peaked feature. We interpret this broadening as an effect of the superposition of different conditions in the ionization states of the plasma and we model the spectra with two thermal components of optically thin plasma in non-equilibrium of ionization (VNEI model in XSPEC) with different temperature, abundances, and ionization time-scales, $\tau$ (this model describes the spectra significantly better than a single-temperature model).
We associate the low $\tau$ component with the ejecta closer to the reverse shock and the high $\tau$ component with the material shocked at earlier stages, i. e. the ejecta at higher distances from the center of the shell.
For each component, we let only the Ca and Fe abundances free to vary and we impose solar abundances for all other elements (but Ni, which is assumed to have the same abundance as Fe). We also add three gaussian components to model the line emission from shocked Ti (if any), Cr, and Mn.
\begin{deluxetable*}{llccccc}
\tabletypesize{\scriptsize}
\tablecaption{Results of the spectral analysis for the regions shown in Fig. \ref{fig:map}.}
\tablewidth{0pt}
\tablehead{
\colhead{Parameter} & \colhead{Region 1} & \colhead{Region 2} & \colhead{Region 3} & \colhead{Region 4} &
\colhead{Region 5}
}
\startdata
$kT_1$ (keV)    & $1.45^{+0.2}_{-0.07}$ & $1.37^{+0.6}_{-0.15}$ & $1.5^{+1.1}_{-0.2}$ & $1.2\pm0.1$ & $1.3^{+0.4}_{-0.2}$ \\
$\tau_1$ ($10^9$ s cm$^{-3}$)& $3.2\pm0.7$ & $3.6\pm1.4$ &  $7^{+3}_{-2.}$ & $3.0\pm1.0$ & $0.6\pm0.3$   \\   
$EM_1$\tablenotemark{*} ($10^{18}$ cm$^{-5}$)& $9^{+19}_{-3}$ & $29^{+25}_{-14}$&$9^{+22}_{-7}$ & $24^{+40}_{-6}$ &  $13^{+0.6}_{-0.5}$ \\
     Ca$_1$               &  $500\pm300$   &  $200\pm140$    &  $600\pm 500$  & $600\pm400$ & $140^{+40}_{-90}$  \\
     Fe$_1$               &  $300\pm200$  &  $70\pm60$  & $200\pm150$ & $200\pm160$    &  $300\pm250$  \\
$kT_2$ (keV)             &    $4.3\pm0.8$     &  $5.0\pm0.9$    & $4.0\pm1.6$ & $3.3^{+1.1}_{-0.4}$  &  $5.0\pm0.4$   \\
$\tau_2$ ($10^9$ s cm$^{-3}$)&     $26\pm2$   & $23\pm3$ & $33^{+18}_{-4}$  &  $27^{+3}_{-4}$  & $25.5^{+1.2}_{-1.0}$  \\   
$EM_2$\tablenotemark{*} ($10^{18} $cm$^{-5}$) &  $8\pm 2$  & $33^{+3}_{-4}$  & $2.5^{+4}_{-1.4}$ &  $13\pm3$  & $10.0^{+0.6}_{-0.9}$  \\
     Ca$_2$              & $21^{+8}_{-4}$   &  $5.3^{+1.3}_{-1.1}$ & $36^{+13}_{-19}$ & $37^{+4}_{-7}$  &  $7.4\pm0.7$ \\
     Fe$_2$              & $21^{+9}_{-4}$ & $2.3^{+0.4}_{-0.3}$ & $30^{+30}_{-18}$  & $28\pm10$   &  $5.4\pm0.4$ \\
    $E_{Ti}$ (keV)       & $4.90^{0.05}_{-0.04}$   &   $4.9$\tablenotemark{**} & $4.93\pm0.06$ & $4.9$\tablenotemark{**} & $4.9$\tablenotemark{**} \\
$N_{Ti}$ ($10^{-6}$ cm$^{-2}/$s)& {\bf2.1$\pm$ 1.2} & $0~(<0.6)$ & {\bf0.5$\pm$0.3} & $0.8^{+0.9}_{-0.8}$  & $0~(<0.5)$ \\
    $E_{Cr}$ (keV)   &$5.50^{+0.02}_{-0.03}$& $5.52\pm0.15$ &  $5.52^{+0.05}_{-0.04}$ & $5.51^{+0.04}_{-0.02}$ & $5.47\pm0.08$  \\
$N_{Cr}$ ($10^{-6}$ cm$^{-2}/$s)& $6.3\pm1.1$  &   $2.3\pm 1.2$   & $0.8^{+0.4}_{-0.3}$ & $3.9\pm0.9$ & $6.1^{+1.5}_{-1.6}$   \\
    $E_{Mn}$ (keV)       &   $6.02\pm0.03$   & $6.04^{+0.04}_{-0.06}$ & $6.1$\tablenotemark{**} & $6.00\pm0.06$ & $6.05\pm0.08$  \\
$N_{Mn}$ ($10^{-6}$cm$^{-2}/$s)&$3.8^{+1.0}_{-1.1}$& $2.0\pm1.1$ & $0.3\pm0.3$ &  $2.8\pm0.8$ & $2.8^{+1.3}_{-1.5}$  \\
Reduced $\chi^2$ (dof)   &   $1.05~(1855)$    &   $0.97~(2022)$  &  $1.01~(203)$  & $0.98~(1399)$ & $1.07~(2521)$ \\  
\enddata
\tablecomments{All errors are at the 90\% confidence level.}
\tablenotetext{*}{Emission measure per unit area. } \tablenotetext{**}{Unconstrained.}
 \label{tab:specres}
\end{deluxetable*}

Table \ref{tab:specres} shows the best-fit parameters for the five spectral regions. The temperatures of the low $\tau$ component (labelled ``1") are significantly lower than those of the high $\tau$ component. This result suggests that the temperature of electrons increases as they move away from the reverse shock, in agreement with what expected in the case of electron heating by Coulomb collisions with hotter protons in the post-shock flow. 

\begin{figure}[tb!]
  \centerline{\hbox{     
      \psfig{figure=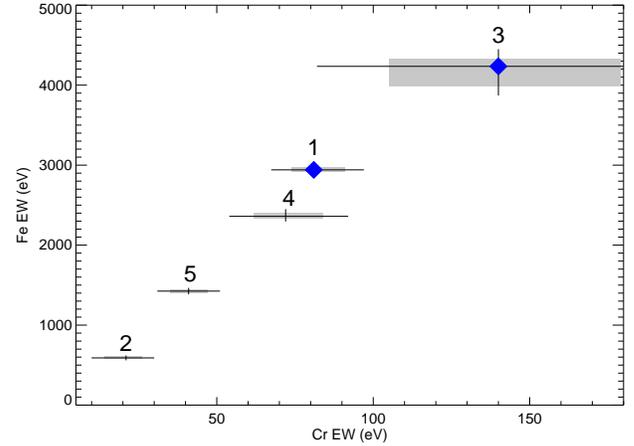,angle=90,width=\columnwidth}}}
 \caption{Fe EW in the $6.1-6.7$ keV energy band versus Cr XXII EW in the $5.3-5.7$ keV band as obtained from the spatially resolved spectral analysis of the spectra extracted from the regions shown in Fig. \ref{fig:map}. Error bars are at the $68\%$ (gray boxes) and $90\%$ (black crosses) confidence levels. The blue diamonds mark the regions where we detected the Ti emission line.}
 \label{fig:ew}
 \end{figure}

Though the Ca and Fe abundances in the low $\tau$ component are poorly constrained, the abundances in the high $\tau$ component show much smaller errors. This is because the high $\tau$ component dominates the flux in the line bands in all the regions, being in the range $\sim63\%-74\%$ of the total in the $3.6-4.05$ keV band, and $65\%-85\%$ of the total in the $6.1-6.7$ keV band.
The pattern of the best-fit abundances shows that the features observed in the EW maps of Fig. \ref{fig:map} are indeed the results of variations in the plasma chemical composition. In particular, region 2 and region 5, characterized by low values of both Ca and Fe EW, show the minimum values of the abundances Ca$_2$ and Fe$_2$ (see Table \ref{tab:specres}). Also, the Fe equivalent width, measured from the spectral analysis in regions $1-5$ (shown in Fig. \ref{fig:ew}) remarkably confirms the results of the EW maps.

Table \ref{tab:specres} shows the presence of Cr and Mn emission lines, whose line centroids are in good agreement with those found by \citet{thn09} in the $Suzaku$ global spectrum of the remnant. Our spatially resolved spectral analysis allows us to find spatial variations in the intensity of these lines. We measured the EW of the Cr and Mn emission lines in all the regions. While the Mn EW presents large error bars, we reveal significant variations in the EW of the Cr XXII line. Figure \ref{fig:ew} shows that the Cr EW derived by our fittings increases in regions with high Fe EW. This correlation strongly indicates that Fe and Cr are spatially co-located, thus confirming what proposed by \citet{ytl13} on the basis of the global spectrum of the remnant.

We also found that the quality of the fits in regions 1 and 3 improves significantly by adding to the model a Gaussian component with energy $E\sim4.90$ keV, corresponding to some transitions to the ground level of Ti \citep{NIST_ASD}. The normalization of this component is higher than zero at almost 3 sigmas ($\Delta\chi^2\sim8$) in region 1 and at more than two sigmas in region 3 ($\Delta\chi^2\sim5$). Though this additional Ti line only affects a few bins of the spectra, it still produces a significant reduction in the $\chi^2$ (which is calculated in the broad $3.6-6.7$ keV energy band), thus indicating that the improvement of the fits in the Ti energy band is large indeed. Figure \ref{fig:spec} shows that the Ti emission line is visible when considering the summed (over all the observations) spectrum of region 1.
Interestingly, we detect significant Ti emission only in the regions with the highest Fe K and Cr equivalent widths (see Fig. \ref{fig:ew}). In regions 2 and 5 (where the Fe and Cr equivalent widths are the lowest) the best-fit value of the normalization of the Ti line is zero. We observe only a marginal ($\sim1.5$ sigma) indication for the Ti line in region 4.
We checked our results by analyzing also the $Chandra$ ACIS spectra of region 1 extracted from observations 10093-97, 10902-4, and 10906 (all performed in 2009, for a total of 734 ks)\footnote{Chandra data have been reprocessed with CIAO 4.7 and spectra have been extracted with the SPECEXTRACT script.}. The combined fits of the pn and ACIS spectra is reasonably good (reduced $\chi^2=1.26$ with 3644 dof), though it presents clear residuals corresponding to the Ca and Fe line complexes, possibly associated with the different PSFs of the two telescopes (a larger contamination from bright nearby regions is expected in the \emph{XMM-Newton} spectra). 
These residuals make the Ti line unconstrained, so we focus on the $4.4-6.1$ keV continuum band, by freezing all the parameters, but the normalization of the Gaussian components, to the best-fit values obtained in the $3.6-6.7$ keV. We obtain a reduced $\chi^2=1.06$ with 1626 dof, and the normalization of the Ti line is $N_{Ti}=1.4^{+0.8}_{-0.7}\times10^{-6}$ cm$^{-2}/$s, in agreement with the value in Table 1. Also, $N_{Ti}>0$ at more than three sigmas.

\section{Discussion and conclusions}

We analyzed several archive \emph{XMM-Newton} observations of Tycho's SNR to study the spatial distribution of the shocked ejecta.
We found strong indications for anisotropies in the distributions of Fe-rich and Ca-rich shocked ejecta, which appear to be mainly localized in the northern part of the remnant. We also found a radial stratification of the ejecta chemical composition, with Ca and Si localized in an outer shell with respect to Fe.

Our spatially resolved analysis shows that the EW of the Cr and Fe lines are correlated in the different regions of the remnant, with regions having the highest Fe abundances showing also the highest Cr equivalent width.
Theoretical models of delayed-detonation show that, besides the explosive Si-burning regime, different yields of Fe-group elements can be synthesized in incomplete Si-burning layers, depending on variations in the details of the transition from deflagration to detonation (e. g., \citealt{ibn99}). The indications of a spatial co-location of Fe and Cr obtained by us suggest that the bulk of shocked Fe-group elements in Tycho's SNR has been synthesized in the explosive Si-burning regime (as in Kepler's SNR, see \citealt{pbm13}).

We also found indications for the presence of Ti line emission, confirmed by the joint $Chandra$ and \emph{XMM-Newton} data analysis. We verified that this emission is concentrated in regions characterized by bright emission from Cr and Fe K. We then conclude that the spatial distribution of the Ti-rich ejecta should follow that of the Fe-rich ejecta in Tycho's SNR.
This clearly suggests that Fe-peak nuclei are spatially co-located in the remnant, in agreement with the predictions of multi-D models of Type Ia SN explosions.

The $Swift/BAT$ observations of radioactive emission from $^{44}$Ti points toward a total mass $M_{44Ti}>10^{-5}M_{\sun}$ \citep{tsl14}. This value is consistent with that expected from a delayed-detonation explosion, which appear to be particularly suited for Tycho's SNR (e.g., \citealt{bbh06}). Delayed-detonation models (e.g., \citealt{ibn99}) generally predict much larger yields of $^{48}$Ti ($5-7\times10^{-4}M_{\sun}$) and $^{50}$Ti ($\sim3\times10^{-4}M_{\sun}$) than $^{44}$Ti. We then expect these heavy isotopes to contribute predominantly to the X-ray line emission.

We can evaluate whether the observed Ti line flux reported in Table 1 is sound by comparing it to the flux of the Cr emission line. In particular, in region 1 we obtain a line flux ratio of the Ti to Cr emission lines $N_{Ti}/N_{Cr}=0.3\pm0.2$ (see Table \ref{tab:specres}).  The expected line flux ratio of the Ti to Cr emission lines is $N_{Ti}/N_{Cr}\sim M_{Ti}/M_{Cr}\times E_{Ti}/E_{Cr}$, where $M_{Ti,Cr}$  indicate the mass of the shocked (i. e., X-ray emitting) Ti and Cr, respectively, and $E_{Ti, Cr}$ are the corresponding emissivities per ion. If we consider that Ti and Cr are spatially co-located, we can assume that the ratio $M_{Ti}/M_{Cr}$ is the same as the ratio of the total (shocked and unshocked) masses of Ti to Cr synthesized at the explosion. 
This mass ratio is predicted to be $M_{Ti}/M_{Cr}\sim 0.06$ \citep{ibn99}. To derive the Ti and Cr emissivities, we adopt the same approach as \citet{bbh08} (see also \citealt{hph00}), by interpolating the K$\alpha$ emissivities of Si, S, Ar, Ca, 
Fe, and Ni for a plasma in non-equilibrium of ionization\footnote{We have used the ATOMDB database and the APEC non-equilibrium ionization library Libapecnei, see \url{http://www.atomdb.org/index.php}.} having the best-fit temperature and ionization timescale that we obtained in region 1 for the high $\tau$ component (see Table \ref{tab:specres}), which is the one that mainly contribute to the line emission. We thus obtain $E_{Ti}/E_{Cr}\sim 2$. Therefore, the expected Ti to Cr line flux ratio is  $N_{Ti}/N_{Cr}\sim 0.12$, which is consistent with that observed in region 1. Though large uncertainties are involved in this estimate, we conclude that the observed Ti flux appears to be reasonable and consistent with expectations.

Further observations are necessary to study in details the spatial distribution of the Ti-rich ejecta in Tycho's SNR. Hard X-ray observations performed with the $NuSTAR$ telescope will allow us to trace the $^{44}$Ti emission with high spatial resolution and to verify if it is consistent with that of the Fe K emission (shown in blue in Fig. \ref{fig:map}), as suggested by our analysis. 
However, we point out that, the abundance of neutron-rich elements is highly sensitive to the electron captures taking place in the central layers of the exploding star, so, in principle, the spatial distribution of $^{48}$Ti and $^{50}$Ti may not coincide with that of the radioactive $^{44}$Ti (though an efficient mixing is expected).

A major leap forward will be provided by the next generation of X-ray telescopes. As an example, we simulated an 80 ks observations of the Fe-rich region of Tycho's SNR, performed with the Soft X-ray Spectrometer of the forthcoming $Astro-H$ mission (\citealt{astro-H14} and references therein).
We verified that it will be possible to detect the Ti line in the whole field of view of the telescope with a very high statistical confidence (5 sigmas). A detailed study of the spatial distribution of the shocked Ti will be possible with the $Athena$ telescope (\citealt{heu13,heu13snr}). We simulated a 50 ks observation of Tycho's SNR performed with $Athena$ X-IFU \citep{rbd14} and found that, by assuming an average line flux equal to that observed in region 1, the X-IFU spectra will allow us to detect the Ti emission line in 1 arcmin$^2$ regions at more than $5\sigma$.




\acknowledgments
We thank the anonymous referee for their comments and suggestions.
This paper was partially funded by the PRIN INAF 2014 grant.
M. M. thanks M. Dadina for discussions about the X-IFU instrumental background. 


\bibliographystyle{apj}




\end{document}